\documentstyle[preprint,aps]{revtex}
\begin{document}
\draft
\title{On the amelioration of quadratic divergences}
\author{R.~Delbourgo\cite{Author1}}
\address{University of Tasmania, GPO Box 252-21, Hobart, 
 Tasmania 7001, Australia}
\author{M.D.~Scadron\cite{Author2}}
\address{University of Arizona, Tucson, AZ 85721, USA}
\date{\today }
\maketitle

\tightenlines

\begin{abstract}
Once massless quadratically divergent tadpole diagrams are discarded, 
because they contain no intrinsic scale, it is possible to convert other 
divergences into logarithmic form, using partial fraction identities; 
this includes the case of quadratic divergences, as has been applied to 
the linear sigma model. However the procedure must be carried out with 
due care, paying great attention to correct numerator factors.
\end{abstract}


\narrowtext

In QED one knows that the formally quadratically divergent photon vacuum
polarization graph is reduced to a logarithmically divergent singularity
by invoking gauge invariance. However, in a linear $\sigma$ model field
theory, involving spinless pion and $\sigma$ mesons, no principle like
gauge invariance can be invoked; yet the quadratic divergence can be tamed
and converted into a logarithmic one. This fact was first noted for a
quark-level SU(2) model, by using the $2\ell$-dimensional regularization 
lemma (DRL) \cite{DS}:
\begin{equation}
I\equiv\int\bar{d}^4p\left[\frac{m^2}{(p^2-m^2)^2}-\frac{1}{p^2-m^2}\right]
 = \lim_{\ell\rightarrow 2} \frac{i(m^2)^{\ell-1}}{(4\pi)^\ell}
   [\Gamma(2-\ell)+\Gamma(1-\ell)] = -\frac{im^2}{16\pi^2},
\end{equation}
since \cite{f1} the combination
$\Gamma(2-\ell)+\Gamma(1-\ell)=\Gamma(3-\ell)/(1-\ell)\rightarrow -1$,
as $\ell\rightarrow 2$. In this note we shall show that lemma (1) holds for
all regularization schemes which set the massless quadratically divergent 
tadpole integral $\int d^4p/p^2$ equal to zero, for the reason that it 
contains no intrinsic mass scale (dimensional analysis). This therefore 
includes dimensional regularization \cite{Le}, analytic and zeta function
regularization, Lodder log \cite{Lo} and Pauli-Villars 
regularizations \cite{f11,DRS}.

To prove the result (1), start with the partial fraction identity
$$\frac{m^2}{(p^2-m^2)^2} - \frac{1}{p^2-m^2} = 
  \frac{1}{p^2}\left[\frac{m^4}{(p^2-m^2)^2}- 1\right], $$
and integrate the rhs over $\int \bar{d}^4p$. Making a Wick rotation to
$q^2 = -p^2, \bar{d}^4p = iq^2 dq^2/16\pi^2$ and dropping the tadpole term, 
we remain with 
\begin{equation}
I = -i\frac{m^4}{16\pi^2}\int_0^\infty \frac{dq^2}{(q^2+m^2)^2} = 
   -\frac{im^2}{16\pi^2}
\end{equation}
coinciding with the rhs of (1). An alternative way of reaching this
conclusion is to apply the Karlson \cite{K} trick (advocated also by
Schwinger):
\begin{equation}
 \frac{d}{dm^2}\int \frac{d^4p}{p^2-m^2} = \int\frac{d^4p}{(p^2-m^2)^2},
\end{equation}
since this serves to eliminate a possible quadratic tadpole infinity.
Thus one may readily verify, without having to make any statement about
quadratic divergences tadpole infinities, that 
$$dI/dm^2=2m^2\int\bar{d}^4p/(p^2-m^2)^3=-i/16\pi^2.$$
The only question/problem is what happens when we integrate this up in $m^2$.
The regularization (1) presumes that no constant $\Lambda^2$ 
intrudes---amounting to setting the tadpole contribution to zero.

At this point we want to inject a note of caution; this concerns numerators
of divergent Feynman integrals. Consider the vacuum polarization integral
\begin{eqnarray}
\Pi^{\mu\nu}(k)&=& ie^2{\rm Tr}\int\frac{[\gamma^\mu((p+k)\cdot\gamma+m)
    \gamma^\nu(p\cdot\gamma+m)]}{((p+k)^2-m^2)(p^2-m^2)} \bar{d}^{2\ell}p
    \nonumber\\
 &=& c\eta^{\mu\nu} + (k^2\eta^{\mu\nu}-k^\mu k^\nu)f(k^2).
\end{eqnarray}
Gauge invariance requires $c=0$ (and dimensional regularization will
guarantee it). Thus whatever the chosen regularization method, we must
insist upon $\Pi^{\mu\nu}(0)$ or $c=0$. In dimensional regularization one 
{\em strictly} finds
\begin{equation}
c_{2\ell}\propto 2^\ell\int\frac{(1/\ell-1)p^2+m^2}{(p^2-m^2)^2}
\bar{d}^{2\ell}p\propto 2^\ell[(1/\ell-1)\Gamma(1-\ell)-\Gamma(2-\ell)/\ell]
 (m^2)^{\ell-1}\equiv 0.
\end{equation}
The delicate cancellation only works because the correct dimensional factors
have been included in the numerator of (4). However if one were blindly to
set $\ell=2$ at the start of eq. (5), one would find
$$c_4 \propto \int \frac{4m^2-2p^2}{(p^2-m^2)^2}\, \bar{d}^4p =
  2\int \frac{\bar{d}^4p}{p^2}\left[-1+\frac{m^4}{(p^2-m^2)^2}\right],$$
so dropping the tadpole term directly in 4-D would leave us with a 
{\em finite} but {\em nonzero} contribution 
$2m^4\int\bar{d}^4p/[p^2(p^2-m^2)^2] \propto m^2$, which is undesirable! 
It is therefore vitally important to incorporate the correct factors in the 
numerators of such integrals, before jumping to conclusions. 
If in any doubt one should apply a regularization scheme which respects the 
symmetries that one holds dear---such as dimensional regularization.

To conclude, we can with appropriate care interpret the regularized 
quadratically divergent integral as
$$16i\pi^2\int\bar{d}^4p/(p^2-m^2)=m^2[\ln(m^2/M^2)+C],$$
where $M$ is some mass scale and $C$ is some indeterminate constant. This
form is of course consistent with the once $m^2$-differentiated integral:
\begin{equation}
-16i\pi^2\left(m^2\frac{d}{dm^2}-1\right)\int\frac{\bar{d}^4p}{p^2-m^2}
= -16i\pi^2I = -m^2,
\end{equation}
and it obeys the DRL, eq (1); but the naive ultraviolet cutoff integral 
$\int d^4p/(p^2-m^2)=\Lambda^2$ violates all decent regularization schemes 
discussed above. 

The fact that dimensional
regularization regards all one-loop divergences as simple poles in $\ell$
means that, with appropriate care, one should be able to convert them all 
into `logarithmic' form by suitable discarding all integrals of the type 
$\int d^4p\,(p^2)^n$, at least for $n>-2$, because they contain no intrinsic
scale. Taming such divergences may have deeper repercussions; in particular,
one of the reasons for invoking the elegant idea of supersymmetry is that it 
provides for a nice cancellation between boson and fermion loop contributions
(including the numbers of each), with consequences for fine-tuning. However 
if these loop contributions are effectively logarithmic, the argument for 
supersymmetric cancellations is weakened considerably and one requires 
instead that $\sum_i(-1)^{F_i} g_i^2m_i^2[\ln m_i^2 + C]$ be finite, rather 
than $\sum_i (-1)^{F_i}g_i^2$, with all $g_i$ normally taken to be equal and
where the summation is over all spin and charge states.

It should be emphasized that we have not actually eliminated the quadratic
divergence, we have simply transmogrified it. [Thus $\lambda\phi^4$ theory
still has a dilatation current anomaly, or anomalous dimension for the 
$\phi$-field, which occurs to order $\lambda^2$ when $\phi$ is massless; but
a massive $\phi$ will contribute to the self-mass renormalization to order
$\lambda$ when the tadpole loop {\em contains a mass}.] Since the above 
results apply to one-loop diagrams one might well ask if the tadpole 
subtraction technique can be extended to higher loops. To show that this 
can be done, we shall exemplify the process by considering Yukawa theory,
$${\cal L}=Z_\phi[(\partial\phi)^2-(\mu^2-\delta\mu^2)\phi^2]/2 +
   Z_\psi\bar{\psi}[i\gamma.\partial -m +\delta m]\psi+
   Z_gg\bar{\psi}\gamma_5\psi\phi +Z_\lambda\lambda\phi^4,$$
focusing on the fermion loop contributions up to order $g^4$ to the meson
self-energy, which are the main source of quadratic infinities. (We have taken
a pseudoscalar model rather than a scalar model because it renders the 
calculations more transparent in the soft meson momentum limit). Specifically
we shall examine the $g^4$ overlapping divergence as a good test case of the
procedure.

Before doing so, let us take note of a couple of one-loop results as they are 
needed at the next perturbative order. The meson self-energy to order
$g^2$ is given by
\begin{equation}
 \Pi^{(2)}(k)=ig^2{\rm Tr}\int\bar{d}^4p\,\gamma_5\frac{1}{\gamma.(p+k)-m}
 \gamma_5\frac{1}{\gamma\cdot p-m}=4ig^2\int\frac{[p\cdot(p+k)-m^2]\,
  \bar{d}^4p}{[(p+k)^2-m^2][p^2-m^2]}.
\end{equation}
For simplicity, renormalise this at $k=0$, allowing the self-mass counterterm
to be massaged from quadratic to logarithmic by tadpole subtraction:
$$\delta\mu^2 = -\Pi^{(2)}(0) = -4ig^2\int\frac{\bar{d}^4p}{p^2-m^2} \rightarrow
   -4ig^2\int\frac{m^2\,\,\bar{d}^4p}{p^2(p^2-m^2)}. $$
The wave-function renormalization $Z_\phi$, being logarithmic, requires
no treatment; nor is it needed for the next part of the argument.
The vertex correction is also logarthmic but {\em is} required to the next order
when considering the overlapping divergent graph, so let us just note that
at zero momentum transfer,
\begin{equation}
\Gamma^{(2)}(p) = ig^2\gamma_5\int \frac{\bar{d}^4p}{(q-p)^2(q^2-m^2)},
\end{equation}
from which it follows that to order $g^2$ one can take
$$Z_g =1-ig^2\int \frac{\bar{d}^4q}{q^2(q^2-m^2)}$$
as the coupling constant counterterm. Observe that we have neglected the mass
of the meson, as that has little bearing on the ultraviolet behaviour: the
difference $1/(k^2-\mu^2)-1/k^2= \mu^2/k^2(k^2-\mu^2)$ has improved high-energy
convergence; non-vanishing $\mu$ just muddies the argument to come.

Now concentrate on the overlapping $g^4$ order fermion loop contribution to
the meson self-energy. {\em Including the vertex counterterm}, it is given by
\begin{eqnarray}
\Pi^{(4)}_{\rm overlap}(0)&=&\!-g^4{\rm Tr}\int\! \bar{d}^4p\left[
    \gamma_5\frac{1}{\gamma\!\cdot\!p\!-\!m}\gamma_5
    \frac{1}{\gamma\!\cdot\!p\!-\!m}\left(\!\int\!
   \frac{\bar{d}^4q}{(q\!-\!p)^2}\gamma_5\frac{1}{\gamma\!\cdot\!q\!-\!m}
   \gamma_5\frac{1}{\gamma\!\cdot\!q-m} +i(Z_g\!-\!1)\right)\right]\nonumber\\
 &=& 4g^4\int\frac{\bar{d}^4p}{p^2-m^2}\int\frac{\bar{d}^4q}{q^2-m^2}
   \left(\frac{1}{q^2}-\frac{1}{(q-p)^2}\right)
   \equiv 4g^4\int\frac{f(p^2)\,\,\bar{d}^4p}{p^2-m^2}.
\end{eqnarray}
The vertex renormalization has guaranteed that the $q$-integral is finite,
but the remaining $p$-integral is potentially quadratically infinite and it is
this part that we wish to convert to logarithmic by tadpole subtraction. The
trick is to notice that $f(p^2)$ can be converted into the dispersive form
$$f(p^2)\propto p^2\int_{m^2}^\infty\frac{ds}{s(s-p^2)}=-\ln
 (1-\frac{p^2}{m^2}),$$
whereupon the overlap integral can be re-expressed as
$$\Pi^{(4)}_{\rm overlap}(0)\propto\int\bar{d}^4p\int ds\left[\frac{1}{s-m^2}
   \left(\frac{1}{p^2-m^2}+\frac{1}{s-p^2}\right)-\frac{1}{s(p^2-m^2)}\right].$$
Having manoeuvred the quadratic infinity into suitable form, we can apply the
amelioration procedure, $\int\bar{d}^4p/(p^2-M^2)\rightarrow \int\bar{d}^4p\, 
M^2/p^2(p^2-M^2)$, and end up with
$$\Pi^{(4)}_{\rm overlap}(0)\propto\int\frac{\bar{d}^4p}{p^2}\int\frac{ds}{s-m^2}
   \left[\frac{m^4}{s(p^2-m^2)} + \frac{s}{s-p^2}\right]. $$
Thus we have achieved our goal of conversion of quadratic to logarithmic in {\em 
both} of the `internal momentum' integrals. This does not imply that the
logarithmic infinities remain at first order like $\log(M^2/m^2)$; indeed
we encounter higher powers of logs in the overlapping example above but the
important point is that the quadratic infinity has been ameliorated. Of course, 
this resulting divergence needs as ever to be subtracted by a new fourth order 
counterterm, but this is just standard renormalization fare and there is no 
problem with that. For higher loop integrals we 
are optimistic that the same procedure should work, provided one replaces the 
logarithmic dispersive integral $f$ by the more general Feynman parametric 
form ($\alpha$ represents a set of Feynman parameters),
$$f(p^2) \sim p^2\int d\alpha/[p^2-M^2(\alpha)],$$
and follows the steps above.


\begin{references}
\bibitem{f1} We define $\bar{d}p$ as $dp/2\pi$ for each momentum
 integration.
\end{references}
\end{document}